\documentstyle[amssymb,aps,preprint]{revtex}

\draft
\tighten
\begin{document}
\title{Effect of Impurities and Effective Masses on Spin-Dependent Electrical
Transport in Ferromagnet-Normal Metal-Ferromagnet Hybrid Junctions}
\author{Zhen-Gang Zhu, Gang Su$^{*}$, Biao Jin and Qing-Rong Zheng}
\address{Department of Physics, The Graduate School of the Chinese Academy of\\
Sciences, P.O. Box 3908, Beijing 100039, China}
\maketitle

\begin{abstract}
The effect of nonmagnetic impurities and the effective masses on the
spin-dependent transport in a ferromagnet-normal metal-ferromagnet junction
is investigated on the basis of a two-band model. Our results show that
impurities and the effective masses of electrons in two ferromagnetic
electrodes have remarkable effects on the behaviors of the conductance,
namely, both affect the oscillating amplitudes, periods, as well as the
positions of the resonant peaks of the conductance considerably. The
impurity tends to suppress the amplitudes of the conductance, and makes the
spin-valve effect less obvious, but under certain conditions the phenomenon
of the so-called impurity-induced resonant tunneling is clearly observed.
The impurity and the effective mass both can lead to nonmonotonous
oscillation of the junction magnetoresistance ($JMR)$ with the incident
energy and the thickness of the normal metal. It is also observed that a
smaller difference of the effective masses of electrons in two ferromagnetic
electrodes would give rise to a larger amplitude of the $JMR$.
\end{abstract}


\section{Introduction}

It is getting realized that the successful control of spin-polarized
electrical currents would give rise to essential impact in information
technology, thereby resulting in that the spin-dependent tunneling in hybrid
systems of ferromagnetic and normal metals (insulators) has attracted much
attention during the past years (see, e.g. Refs.\cite{as,gi,prinz,mood} for
review). A nascent field dubbed as {\it spintronics, }is therefore emerging.
Among others, a large number of studies on the electrical transport
properties in metallic magnetic multilayers show the existence of the effect
of a so-called giant magnetoresistance (GMR)\cite{gmr}, which is now
believed to be caused by spin-dependent scattering in the systems. Since its
potential applications, the GMR effect has been actively studied both
theoretically and experimentally in recent years (see, e.g. Refs.\cite
{gi,svalve}).

Apart from the GMR found in metallic magnetic multilayers, the spin-valve
effect is observed in ferromagnetic (FM) tunneling junctions. Following the
discovery of spin-polarized tunneling of conduction electrons in
ferromagnets by Tedrow and Meservey\cite{ted}, Julli\`{e}re was the first to
observe the effect that the tunneling resistance in an F-I-F junction
consisting of two FM electrodes separated by an insulating (I) barrier
depends strongly on the relative orientations of magnetizations, namely the
resistance is low when magnetizations in FM electrodes are parallel, and the
resistance is high when magnetizations in FM electrodes are antiparallel,
and a two-current model was proposed to interpret the observed large
junction magnetoresistance in Fe-Ge-Co junctions at $4.2$ $K$ by assuming
the spin conservation so that the tunneling of spin up and spin down
electrons undergoes two independent processes\cite{julliere}. The tunneling
magnetoresistance has also been observed by Maekawa and G\"{a}fvert\cite
{maeka}. Treating the F-I-F junction as a single quantum-mechanical system,
Slonczewski adopted a free-electron model to describe the tunneling of
spin-polarized conduction electrons, and recovered the spin-valve effect\cite
{slonczewski}. The extension of Slonczewski's model to a realistic band
structure based on a tight-binding model was established\cite{mathon}.
Recently, on the basis of scattering matrix theory the spin-dependent
electrical and thermal transport in F-I-F junctions was developed at finite
bias voltage and at finite temperatures\cite{wang-su}. It is found that not
only the electrical conductance but also the thermal conductance reveals the
spin-valve effect at low temperatures. There are also some theories devoted
to F-I-F tunneling junctions (see Ref.\cite{mood} for a review). Moreover,
the spin-valve effect can also appear in ferromagnet-normal
metal-ferromagnet (F-N-F) junctions which may have different transport
properties from the F-I-F junctions. Johnson was the first to predict the
phenomenon of spin accumulation in F-N-F junctions, which was soon confirmed
in a three-terminal device, and a transistor effect was observed in such a
device\cite{johnson}.\ Valet and Fert studied the F-N-F structure and
applied the Boltzmann equation to the current perpendicular to the plane
structure by assuming that the thickness of nonmagnetic metallic layer is
much shorter than the spin-diffusion length. The theory can reduce to
Johnson's theory as the thickness of nonmagnetic metal is much longer than
the spin-diffusion length\cite{Valet}. By taking into account different
magnetizations of ferromagnetic reservoirs and spin diffusion in the normal
metal, Hernando et al studied the equation for the diffusive spin transport
in the F-N-F structure and showed that the difference between the
conductance of the parallel and antiparallel configuration can be either
positive or negative as a function of the applied magnetic field\cite
{Hernando}. Zheng et al have investigated a double tunnel junction model
describing the F-I-N-I-F structure by means of quantum mechanical approach,
and the oscillation of the tunneling conductance and the TMR on the spacer's
thickness is found\cite{zheng}. A finite-element method\cite{brat} and
nonequilibrium Green function method\cite{green} has also been applied to
the mesoscopic junction systems.

Within the theories above mentioned, the effect of impurities was not much
investigated. Although it is argued that the impurities and defects may lead
to the resonant tunneling in F-I-F junctions\cite{bratkov}, the study on the
effect of impurities on the spin-dependent transport in F-N-F junctions is
still sparse. Another important factor, which is usually ignored in
theories, is the effect of the effective masses of electrons in FM and
normal layers. As in realistic hybrid junctions the electrodes are made by
different ferromagnets, such as half-metals \cite{bratkov} and magnetic
semiconductors\cite{ohno1,ohno2,fla}, the effective mass would have
significant effect on the transport properties. In this paper, by treating
the F-N-F junction as a quantum-mechanical system we shall investigate the
effect of nonmagnetic impurities as well as the effective masses of
electrons in FM and normal layers on the spin-dependent electrical transport
properties in spin-valve systems. It is found that the impurity and the
effective mass can both lead to nonmonotonous oscillation of $JMR$ with the
incident energy and the thickness of the normal metal. The effective mass
can enlarge the $JMR$ effect when the difference between the two FM layers
is small. On the other hand, if the difference is large, the $JMR$ effect
becomes smaller and the inverse $JMR$ effect would appear.

The rest of this paper is outlined as follows. In Sec. II the model which we
shall adopt in the subsequent sections will be described and some relevant
formalism will be derived. Sec. III will present the effect of nonmagnetic
impurities on the junction magnetoresistance in F-N-F junctions, and the
effect of effective masses of electrons in FM electrodes will be given in
Sec. IV. Finally, a summary will be presented.

\section{Model and Formalism}

Consider an F-N-F junction consisting of two ferromagnetic layers separated
by a normal metal layer with thickness $d$. In a nearly-free-electron
approximation of the spin-polarized conduction electrons, the longitudinal
part of the effective one-electron Hamiltonian can be written as

\begin{equation}
H=-\frac{\hbar ^{2}}{2m_{j}^{\ast }}\frac{d^{2}}{dx^{2}}+U(x)+[\gamma
_{1}\delta (x)+\gamma _{2}\delta (x-d)]-{\bf h}(j)\cdot {\bf \sigma }
\label{a}
\end{equation}
where $m_{j}^{\ast }$ ($j=1,2,3$) is the effective mass of an electron in
three regions (Region 1: left FM electrode; Region 2: normal metal; Region
3: right FM electrode), respectively. The potential $U(x)$ is composed of
three parts: a small bias voltage $V$, two contact potentials $V_{1}$ and $%
V_{2}$ defined as $U(x)=V_{1}$ for $x\leq 0$, $U(x)=V_{2}$ for $x\geq d$ and
zero otherwise. The contact potentials between the ferromagnetic layers and
the normal metallic layer could be effectively considered to be caused
either by roughness of interfaces or by two equivalent, thin insulating
layers. ${\bf h}(j)$ represents the molecular field in FM electrodes, and $%
{\bf \sigma }$ is the conventional Pauli matrices. By assumption, ${\bf h}=0$
inside the nonmagnetic normal metal. In FM electrodes ${\bf h=h}_{L}$ or $%
{\bf h}_{R}$ is constant with relative orientation labeled by angle $\theta $%
. The potentials induced by impurities or defects on the interfaces between
the ferromagnetic and nonmagnetic metal layers have been discussed in Ref. 
\cite{levy}. Generally speaking, the delta-potentials of impurities or
defects on interfaces depend on three-dimensional coordinates. To simplify
the problem and to get useful analytical results, the potentials are
supposed to have forms of $\delta $ potentials as shown in Eq. (\ref{a})
where $\gamma _{1}$ and $\gamma _{2}$ are corresponding coefficients. We
expect that such an approximation would not cause significate qualitative
changes on the interested transport properties, as the motion of electrons
along the transverse direction are not emphasized in the present approach.
The schematic layout of the F-N-F junction is depicted in Fig. 1. Inside the
ferromagnets, a two-band model will be used to describe the motion of
spin-polarized electrons. As spin up and spin down bands are split by
exchange interactions, the electrons with spin up and down have different
energies in FM electrodes, and the one-electron energy is $E_{1x}=\hslash
^{2}k_{1\sigma }^{2}/2m_{1}^{\ast }-\sigma h_{L}$ $+$ $V_{1}$ for the left
FM electrode, and $E_{3x}=\hslash ^{2}k_{3\sigma }^{2}/2m_{3}^{\ast }-\sigma
h_{R}$ $+$ $V_{2}$ for the right FM electrode with $\sigma =\pm 1$
corresponding to $\sigma =\uparrow ,\downarrow $ respectively. Inside the
nonmagnetic normal metal layer, the one-electron energy is $E_{2x}=\hslash
^{2}k^{2}/2m_{2}^{\ast }$.

Consider an incident plane wave of spin-up electrons with unit flux in
Region 1. The solution of the Schr\"{o}dinger equation with eigenvalue $%
E_{1x}$ in the region of the left FM electrode can be written as

\begin{equation}
\psi _{1\uparrow }=k_{1\uparrow }^{-1/2}e^{ik_{1\uparrow }x}+R_{\uparrow
}e^{-ik_{1\uparrow }x},\text{ \ \ }\psi _{1\downarrow }=R_{\downarrow
}e^{-ik_{1\downarrow }x}.  \label{w1}
\end{equation}
In the normal metal region, the eigenfunctions of $H$ with eigenvalue $%
E_{2x} $ are written as

\begin{equation}
\psi _{2\sigma }=A_{\sigma }e^{ikx}+B_{\sigma }e^{-ikx},\text{ \ }\sigma
=\uparrow ,\downarrow .  \label{w2}
\end{equation}
In the region of the right FM electrode, the eigenfunctions of $H$ with
eigenvalue $E_{3x}$ have the form

\begin{equation}
\psi _{3\sigma }=C_{\sigma }e^{ik_{3\sigma }(x-d)},\text{ \ }\sigma
=\uparrow ,\downarrow .  \label{w3}
\end{equation}
The coefficients $R_{\uparrow }$, $R_{\downarrow }$, $A_{\sigma },B_{\sigma
},C_{\sigma }(\sigma =\pm 1)$ are to be determined by properly matching $%
\psi _{j\sigma }$ and $d\psi _{j\sigma }/dx$ at the interfaces $x=0$ and $%
x=d $ \ (the boundary conditions). Owing to the spin conservation the spin
directions are fixed as the electrons move from Region1 to Region 2. The
existence of $\delta $-like impurity potential at the interfaces makes the
derivatives of wave functions no longer continuous. For the simplicity we
shall set $m_{1}^{\ast }=m_{2}^{\ast }=m_{3}^{\ast }=m$ first, and then
consider the effect of different effective masses in Sec. IV. At the
interface $x=0$, the boundary condition is

\begin{equation}
\psi _{1}(0)=\psi _{2}(0),\text{ \ }\frac{d\psi _{2}(0)}{dx}-\frac{d\psi
_{1}(0)}{dx}=\mu _{1}\psi _{2}(0),  \label{ss}
\end{equation}
where $\mu _{1}=\frac{2m_{2}^{\ast }\gamma _{1}}{\hbar ^{2}}.$ At the
interface $x=d$, the boundary condition is

\begin{equation}
\psi _{2}(d)={\frak R}\psi _{3}(d),\text{ \ }\frac{d\psi _{2}(d)}{dx}={\frak %
R}[\frac{d\psi _{3}(d)}{dx}-\mu _{2}\psi _{3}(d)],  \label{bbb}
\end{equation}
where the rotation matrix ${\frak R}=\left( 
\begin{array}{cc}
\cos \frac{\theta }{2} & \sin \frac{\theta }{2} \\ 
-\sin \frac{\theta }{2} & \cos \frac{\theta }{2}
\end{array}
\right) $ and $\mu _{2}=\frac{2m_{3}^{\ast }\gamma _{2}}{\hbar ^{2}}$. In
accordance with Eqs. (\ref{w1})-(\ref{bbb}), all unknown coefficients
introduced above can be obtained. The results are collected in Appendix,
where the quantities $q_{1}(\equiv \frac{m_{2}^{\ast }}{m_{1}^{\ast }}%
)=q_{2}(\equiv \frac{m_{3}^{\ast }}{m_{2}^{\ast }})=1$ should be taken due
to the assumption of the same effective masses here. (The case $q_{1}\neq
q_{2}$ will be discussed later.) In comparison to the results in F-I-F
junctions as discussed by Slonczewski\cite{slonczewski}, our equations (\ref
{w2}), (\ref{ss}) and (\ref{bbb}) for F-N-F junctions with impurities are
quite different. It is these differences that enable the transport
properties in two kind of junctions to show different behaviors.

The transmissivity $T_{p\sigma }=%
\mathop{\rm Im}%
\sum_{\sigma }\psi _{\sigma }^{\ast }(d\psi _{\sigma }/dx)$ can be derived
analytically

\begin{equation}
T_{p\uparrow }=\frac{R_{\downarrow }^{\ast }R_{\downarrow }}{%
4(a_{3}-a_{4})^{2}}[\frac{k_{3\uparrow }(c_{31}^{2}+c_{32}^{2})}{\sin ^{2}%
\frac{\theta }{2}}+\frac{k_{3\downarrow }(c_{61}^{2}+c_{62}^{2})}{\cos ^{2}%
\frac{\theta }{2}}],  \label{dc}
\end{equation}
where $R_{\downarrow },a_{3},a_{4},c_{31},c_{32},c_{61},c_{62}$ are given in
the Appendix. $T_{p\uparrow }$ is valid for spin-up incident electrons and $%
T_{p\downarrow }$ is valid for spin-down electrons by the same expression as
Eq.(\ref{dc}) with $k_{1\uparrow }$ and $k_{1\downarrow }$, $k_{3\uparrow }$
and $k_{3\downarrow }$ interchanged. Thus the total transmissivity in the
two-band case, $T_{p}(\theta )=T_{p\uparrow }(\theta )+T_{p\downarrow
}(\theta ),$ may be obtained.

At zero temperature, it can be reasonably assumed that the electrons with $%
E_{x}$ near Fermi energy $E_{F}$ carry most of the current. With this
assumption and by summing the charge transmission over $E_{x}$ and ${\bf k}%
_{\Vert }$ for the occupied states in the usual manner\cite{duke}, one can
find the conventional expression of surface conductance at a small bias
voltage

\begin{equation}
G(\theta )=\frac{I_{e}}{V}=\frac{m_{1}^{\ast }e^{2}}{2\pi ^{2}\hbar ^{3}}%
E_{F}T_{p}(\theta ),  \label{cc}
\end{equation}
where $m_{1}^{\ast }=m_{2}^{\ast }=m_{3}^{\ast }=m.$ The junction
magnetoresistance (JMR) is defined as usual 
\begin{equation}
JMR(\theta )=\frac{G(\theta =0)-G(\theta )}{G(\theta =0)},  \label{v}
\end{equation}
which reflects the relative change of the conductance with respect to the
different orientation of magnetic moments in FM electrodes, an quantity
being measurable experimentally. Within the framework of the above
formalism, one may investigate the effect of nonmagnetic impurities and the
effective masses of conduction electrons on the conductance in F-N-F
junctions.

\section{Effect of nonmagnetic impurities}

To investigate the effect of nonmagnetic impurities on the conductance of
F-N-F junctions, we have to make proper assumptions. First, we take $%
m_{1}^{\ast }=m_{2}^{\ast }=m_{3}^{\ast }=m$, and suppose the two FM
electrodes the same, i.e. $\left| {\bf h}_{L}\right| =\left| {\bf h}%
_{R}\right| =h$. Then, without loss of generality the coefficients of $%
\delta $-like impurity potentials at the interfaces are taken as $\gamma
_{1}=\gamma _{2}=\gamma $. We have found that $\gamma _{1}\neq \gamma _{2}$
does not alter qualitatively the behaviors observed. The Fermi energy of the
normal metal layer is assumed to be the same as in FM electrodes.

We find that the conductance depends strongly on the relative orientation of
magnetizations of FM electrodes, as shown in Fig.2. It is observed that for
given $d$ (the thickness of the normal metal layer) and $\gamma $ (the
amplitude of the impurity potential), the conductance is the largest at $%
\theta =0$, while at $\theta =\pi $ the conductance reaches the smallest,
resurging the spin-valve effect. However, one may note that the conductance
is not zero as $\theta =\pi $, indicating the spin-valve effect is
imperfect. This is because in the two-band model the spins in the left FM
electrode are not perfectly polarized, there is a portion of electrons with
spin down at the Fermi level available, leading to a small current flowing
through the junction even if the magnetizations in two FM electrodes are
antiparallel, a result quite close to the realistic situation. With
increasing $\theta $ to $\pi $ the conductance $G(\theta )$ is monotonically
decreasing, while $JMR(\theta )$ is monotonically increasing. In
experiments, people are usually interested in the quantity $JMR\equiv
JMR(\theta =\pi )$. Without particular specification the $JMR$ refers to $%
JMR(\pi )$ hereafter.

The effect of impurity on the conductance is shown in Fig.3 for $\theta =\pi
/3$ for an example. One may see that at a given Fermi energy (e.g. $%
E_{F}=2.5eV$) with increasing $\gamma $ the conductance $G$ is first
decreasing to a minimum, then increasing to a maximum, and then decreasing
to zero. When the energy barrier produced by impurities is high enough, the
conductance tends to zero, which is quite reasonable in physics, because in
this situation the electrons cannot overcome the impurity-barrier to tunnel
through the junction. It is interesting to note that $G$ has a maximum at a
certain value of $\gamma $, suggesting that the resonant tunneling occurs in
this case. This could be viewed as a kind of the impurity-induced resonant
tunneling. As the impurity potentials in our model are simulated by the
double-$\delta $ potential barriers, in the region of normal metal layer
there might be some quasi-bound states under certain conditions. The
appearance of the quasi-bound states is a consequence arised from both the
double-$\delta $ potential barriers at two interfaces and the quantum size
effect of the middle metal film. When the incident energy of electrons
matches approximately the level of the quasi-bound states, the resonant
tunneling occurs, manifested by the occurrence of the peak of the
conductance versus $\gamma $. For other values of $E_{F}$, the conductance
has the similar behaviors, in spite of positions of maxima being different.
From Fig. 3, one may observe that the conductance is oscillating with the
Fermi energy.

We also study the $JMR$ versus $\gamma $, as shown in Fig.4. It is seen that
the $JMR$ is nonmonotonous increasing with $\gamma $. At low $E_{F}$ (e.g. $%
2.4eV$) the $JMR$ shows one peak, and with increasing $E_{F}$ (e.g. $2.6eV$)
the $JMR$ has two peaks. When $E_{F}$ takes a larger value (e.g. $2.7eV$)
only one peak of the $JMR$ is left, which is a consequence of the
conductance oscillating with the Fermi energy, leading to that the resonant
positions of the $JMR$ depends strongly on the magnitudes of Fermi energies.
At larger $\gamma $, the $JMR$ saturates to a constant. Furthermore, one may
observe that the $JMR$ is negative for some low $\gamma $ and low $E_{F}$.
This is not difficult to understand by noting the fact that the spin-up and
spin-down electrons in Region 1 have the same incident energies but with
different wave vectors, and the transmissivity $T_{p\sigma }(\theta )$ is
oscillatory increasing with increasing $E_{F}$. At some regimes of low $%
\gamma $ and $E_{F}$, it is found that $T_{p\uparrow }(0)-T_{p\uparrow }(\pi
)>0$ but $T_{p\downarrow }(0)-T_{p\downarrow }(\pi )<0$, while the absolute
value of the latter is greater than the former, resulting in that the
conductance $G$ of parallel alignment of magnetizations is smaller than that
of antiparallel alignment at some low $E_{F}$ and $\gamma $, which finally
gives rise to the negative $JMR$. This is actually a result of the two-band
model. The $JMR$ is oscillating with the Fermi energy, as shown in Fig.5.
When $\gamma =0$, the $JMR$ is symmetrically oscillating with respect to
zero axis. This can be readily understood by taking the two-band feature
into account. With increasing $\gamma $, the $JMR$ becomes asymmetrically
oscillating with more oscillating peaks. Such behaviors are closely related
to the quasi-bound states induced by impurity barriers. If $E_{f}$ is very
large, the spin-dependent effect will be smaller, leading to the $JMR$
approaching to zero with increasing $E_{f}$.

The dependence of the $JMR$ on the thickness of the normal metal layer is
depicted in Fig. 6. It can be seen that the $JMR$ is oscillating with the
thickness $d$. In absence of the impurity potential, the $JMR$ is
symmetrically oscillating with increasing the thickness with one oscillation
period. This phenomenon can be of quantum origin, and more specifically, of
the quantum-size effect, which has been noted before for the GMR in a
one-band model\cite{Barnas} and in double tunnel junction structure\cite
{zheng}. With increasing $\gamma $, the oscillation period becomes two,
which is closely related to the impurity-induced resonant tunneling or the
quasi-bound states. It is interesting to note that the positive and negative 
$JMR$ effect was also found in the giant magnetoresistance of F-N-F system 
\cite{Barnas}, as well as in N/F/I/F/N structure\cite{zhang}.

\section{Effect of Effective Masses}

Now we turn to consider the case that the electrons in the three layers have
different effective masses. The solutions of Schr\"{o}dinger equation have
the same forms as Eqs. (\ref{w1}) - (\ref{w3}). However, the boundary
conditions are changed. At the interface $x=0$, it is

\begin{equation}
\psi _{1}(0)=\psi _{2}(0);\text{ }\frac{d\psi _{2}(0)}{dx}-q_{1}\frac{d\psi
_{1}(0)}{dx}=\mu _{1}\psi _{2}(0),  \label{boundary}
\end{equation}
and at the interface $x=d,$ it becomes

\begin{equation}
\psi _{2}(d)=\Re \psi _{3}(d);\text{ }q_{2}\frac{d\psi _{2}(d)}{dx}=\Re
\lbrack \frac{d\psi _{3}(d)}{dx}-\mu _{2}\psi _{3}(d)].  \label{bccc}
\end{equation}
Recall that $q_{1}=\frac{m_{2}^{\ast }}{m_{1}^{\ast }}$, $q_{2}=\frac{%
m_{3}^{\ast }}{m_{2}^{\ast }}$, $\mu _{1}=\frac{2m_{2}^{\ast }\gamma _{1}}{%
\hbar ^{2}}$, and $\mu _{2}=\frac{2m_{3}^{\ast }\gamma _{2}}{\hbar ^{2}}$.
Note that $q_{1}\neq q_{2}$ and $\mu _{1}\neq \mu _{2}$ here. The
coefficients contained in the wave functions are given in the Appendix. As
the effective masses are different in three regions, the transmissivity,
according to its definition $T_{p\uparrow }=j_{3}/j_{1}=\frac{m_{1}^{\ast }}{%
m_{3}^{\ast }}(%
\mathop{\rm Im}%
\sum_{\sigma }\psi _{\sigma }^{\ast }(d\psi _{\sigma }/dx))$, has the form
of 
\begin{equation}
T_{p\uparrow }=\frac{m_{1}^{\ast }}{m_{3}^{\ast }}\frac{R_{\downarrow
}^{\ast }(\theta )R_{\downarrow }(\theta )}{4(a_{3}-a_{4})^{2}}[\frac{%
k_{3\uparrow }(c_{31}^{2}+c_{32}^{2})}{\sin ^{2}\frac{\theta }{2}}+\frac{%
k_{3\downarrow }(c_{61}^{2}+c_{62}^{2})}{\cos ^{2}\frac{\theta }{2}}].
\label{Tp-up}
\end{equation}
$T_{p\downarrow }$ has the same expression as Eq.(\ref{Tp-up}) but with $%
k_{1\uparrow }$ and $k_{1\downarrow }$, $k_{3\uparrow }$ and $k_{3\downarrow
}$ interchanged. The total transmissivity in the two-band case is still $%
T_{p}(\theta )=T_{p\uparrow }(\theta )+T_{p\downarrow }(\theta )$. The
conductance $G(\theta )$ and the $JMR$ are also defined as Eqs. (\ref{cc})
and (\ref{v}). To investigate the effect of the effective masses of
electrons on the transport properties in F-N-F junctions, in the following
we shall assume the same values of parameters (except for $q_{1}\neq q_{2}$
and $\mu _{1}\neq \mu _{2}$) as those in Sec. III. In this case, the
wavevectors are different for different effective masses.

The $\gamma $- and $E_{F}$-dependences of the conductance $G$ for different
effective masses are plotted in Figs. 7 (a) and (b), respectively, where we
have taken $m_{2}^{\ast }$ as a scale for effective masses. It is seen that
compared with Fig. 3 the effect of the effective masses on the conductance
versus $\gamma $ and $E_{F}$ is quite dramatic. From Fig. 7(a) one may
observe that for a given $m_{1}^{\ast }$ the conductance $G$ versus $\gamma $
is monotonous decreasing for smaller $m_{3}^{\ast }$ but shows resonant
peaks for larger $m_{3}^{\ast }$, namely, when the ratio $m_{1}^{\ast
}/m_{3}^{\ast }$ is larger, the conductance versus $\gamma $ shows no
resonant peaks, implying that the effect of the effective masses smears out
the impurity-induced resonant tunneling, while $m_{1}^{\ast }/m_{3}^{\ast }$
is near unity, the sharp peak appears in the curve of $G$ {\it vs}. $\gamma $%
, and the impurity-induced tunneling recovers. This observation might be
related to the quasi-bound states induced by impurities. The conductance $G$
versus the Fermi energy is oscillating, but the behavior is quite different
for different effective masses, as shown in Fig. 7(b). There are two notable
characters about $G$ varying with the Fermi energy, namely, the oscillating
amplitudes of $G$ for $m_{1}^{\ast }/m_{3}^{\ast }=1$ are always larger than
those for $m_{1}^{\ast }/m_{3}^{\ast }\neq 1$, and the conductance is
increasingly oscillating with increasing the Fermi energy.

In order to figure out the effect of the effective masses on the $JMR$, we
plot the $JMR$ versus $m_{3}^{\ast }$ for different $m_{1}^{\ast }$'s, as
shown in Fig. 8, where we have taken $\gamma =0$ for simplicity. It is seen
that the $JMR$ exhibits maxima when the condition $m_{1}^{\ast }\approx
m_{3}^{\ast }$ is satisfied, which is clearly illustrated in the inset of
Fig. 8, where the contour plot for the maximum $JMR$ in the $m_{1}^{\ast }$-$%
m_{3}^{\ast }$ plane is presented. Away from this condition, the JMR is
sharply increasing for smaller $m_{3}^{\ast }$ and decreasing for larger $%
m_{3}^{\ast }$. This result shows that to get larger $JMR$ one may do the
best to choose those ferromagnetic metals with almost the same effective
masses of electrons to fabricate the two FM electrodes in a spin-valve. The
smaller the difference of the effective masses in the two FM electrodes has,
the larger the $JMR$ is.

We have also studied the thickness of the normal metal dependence of the $%
JMR $ for different $m_{1}^{\ast }$ and $m_{3}^{\ast }$, as depicted in Fig.
9, where we have taken $\gamma =0$. In comparison to Fig. 6, one may find
that different effective masses would lead to different oscillating
behaviors of the $JMR$. In the upper panel of Fig. 9 where $m_{1}=0.05$,
with increasing $m_{3}^{\ast }$ the periodic oscillations of the $JMR$ with $%
k_{F}d$ becomes quite different, and when $m_{3}^{\ast }>1$ the $JMR$
becomes negative. In the middle panel of Fig. 9 where $m_{1}^{\ast }=1$, one
may find that except $m_{3}^{\ast }\approx m_{1}^{\ast }$ the oscillation
amplitudes are much suppressed as the ratio $m_{1}^{\ast }/m_{3}^{\ast }$ is
much larger or smaller than one. In this case the $JMR$ is negative when $%
m_{3}^{\ast }$ is smaller. In the lower panel of Fig. 9 where $m_{1}=10$,
the situation is just opposite to the case in the upper panel, namely, with
increasing $m_{3}^{\ast }$ the oscillating $JMR$ becomes positive from
negative. It appears that the sign of the $JMR$ is primarily controlled by $%
m_{1}^{\ast }$. As mentioned above, the cause for the oscillation of the $%
JMR $ with the thickness of the normal metal layer stems from the
quantum-size effect, while the oscillation period and amplitude are
remarkably affected by the effective masses of electrons, i.e. $m_{1}^{\ast
} $ and $m_{3}^{\ast } $.

To get the knowledge of the resultant effect of both impurity and effective
masses on the $JMR$, we present the results shown in Figs. 10. It is
observed that for a given $m_{1}^{\ast }=0.05$ the $JMR$ is increasing to a
maximum due to the impurity-induced tunneling, then decreasing to a minimum,
and then increasing to saturate with increasing $\gamma $, and those maxima
and minima are decreasing with increasing $m_{3}^{\ast }$, as shown in Fig.
10 (a). With slightly increasing $m_{1}^{\ast }$, the behaviors of the $JMR$
versus $\gamma $ and $m_{3}^{\ast }$ are not altered so much, but the
resonant positions are changed for different $m_{1}^{\ast }$'s, as presented
in Fig. 10 (b). When $m_{1}^{\ast }$ is larger (e.g. $m_{1}^{\ast }=5$), the
situation changes. Though the behaviors of the $JMR$ versus $\gamma $ are
qualitatively similar for $m_{3}^{\ast }<1$, two dips appear for $%
m_{3}^{\ast }>1$, and those maxima and minima are increasing with increasing 
$m_{3}^{\ast }$, as given in Fig. 10 (c). It is indicated that even in the
presence of impurities the difference of the effective masses between the
left and the right FM electrodes would yield dramatic effect on the
behaviors of the $JMR$. To obtain larger $JMR$, the proper ratio $%
m_{1}^{\ast }/m_{3}^{\ast }$ should be carefully chosen.

\section{Summary}

Based on a two-band model in this paper, we have investigated the effects of
nonmagnetic impurities and effective masses of electrons in FM and normal
metallic layers on the spin-dependent transport in a F-N-F hybrid structure
by treating the system as a quantum-mechanical system. It is observed that
both impurities and effective masses have remarkable effects on the
conductance and the $JMR$ as well. The spin-valve effect, though imperfect
due to the two-band feature, is recovered for low impurity barriers, while
it becomes less obvious for high barriers. The so-called impurity-induced
resonant tunneling is clearly seen in the F-N-F system. The reason for this
property is that the quasi-bound states induced by impurities are formed in
this system, and if the incident energy meets with the quasi-bound energy,
the resonant tunneling occurs, leading to peaks observed in the curve of the
conductance versus the amplitude of the impurity potential. It is found that
the $JMR$ is oscillating with the amplitude of the impurity potential, the
incident energies of electrons, as well as the thickness of the normal
metallic layer. The impurity is also observed to influence remarkably the
oscillating amplitudes and periods of the $JMR$. The effective masses of
electrons in the three layers were found to have significant effects on the
conductance and the $JMR$. It is demonstrated that the difference between
the effective masses would tend to suppress the amplitudes of the
conductance, namely, the larger the difference, the smaller the amplitude of
the conductance. In other words, a smaller difference of the effective
masses in the two FM electrodes would give rise to a larger amplitude of the 
$JMR$, suggesting that to get larger $JMR$ one ought ot choose those FM
materials with almost identical effective masses of electrons. We have also
found that the effective masses affect considerably the oscillating periods,
amplitudes, as well as the positions of resonant peaks of the $JMR$.
Recently magnetic semiconductors are used to be the injector of electrons in
both resonant tunneling diodes\cite{ohno1} and light-emitting diodes\cite
{ohno2}. Magnetic semiconductor has both properties of semiconductor and
ferromagnet, and in these systems the effective mass of conduction electrons
would have considerable effect on the transport properties. Therefore, our
above discussion would be meaningful, and the reported results could be
examined experimentally. Here we would like to point out that although we do
not incorporate factors like spin-flip scatterings from magnons and magnetic
impurities, as well as the spin accumulation, our obtained results are
expected to shed some useful light on the spin-dependent transport in F-N-F
hybrid junctions. How to develop a microscopic theory which includes
properly the effects of correlations between electrons, is a challenging
issue, which is now in progress.

\section*{Acknowledgments}

We thank Dr. Z.C. Wang for useful discussions. This work is supported in
part by the National Natural Science Foundation of China (Grant No.
90103023, 10104015), the State Key Project for Fundamental Research in
China, and by the Chinese Academy of Sciences.

\section*{Appendix}

In this Appendix, the coefficients introduced in solving the Schr\"{o}dinger
equations in Secs. II and IV are collected here without giving detail
derivations.

\[
R_{\uparrow }=k_{1\uparrow }^{-1/2}\frac{c_{3}c_{4}+c_{1}c_{6}\tan ^{2}\frac{%
\theta }{2}}{c_{3}c_{5}+c_{2}c_{6}\tan ^{2}\frac{\theta }{2}},\text{ \ }%
R_{\downarrow }=k_{1\uparrow }^{-1/2}\tan \frac{\theta }{2}\frac{%
c_{2}c_{4}-c_{1}c_{5}}{c_{3}c_{5}+c_{2}c_{6}\tan ^{2}\frac{\theta }{2}}; 
\]

\[
A_{\uparrow }=\frac{1}{2}[(a_{1}^{\ast }+1)k_{1\uparrow
}^{-1/2}-(a_{1}-1)R_{\uparrow }],\text{ \ }A_{\downarrow }=\frac{1}{2}%
(1-a_{2})R_{\downarrow }, 
\]

\[
B_{\uparrow }=\frac{1}{2}[(1-a_{1}^{\ast })k_{1\uparrow
}^{-1/2}+(a_{1}+1)R_{\uparrow }],\text{ \ }B_{\downarrow }=\frac{1}{2}%
(1+a_{2})R_{\downarrow }; 
\]
\[
C_{\uparrow }=\frac{c_{3}R_{\downarrow }}{2(a_{3}-a_{4})\sin \frac{\theta }{2%
}},\text{ \ }C_{\downarrow }=\frac{c_{6}R_{\downarrow }}{2(a_{3}-a_{4})\cos 
\frac{\theta }{2}}\text{ }; 
\]
where

\begin{eqnarray*}
c_{1} &=&c_{11}+ic_{12},\text{ }c_{2}=c_{21}+ic_{22},\text{ }%
c_{3}=c_{31}+ic_{32}, \\
\text{ }c_{4} &=&c_{41}+ic_{42},\text{ }c_{5}=c_{51}+ic_{52},\text{ }%
c_{6}=c_{61}+ic_{62},
\end{eqnarray*}

\begin{eqnarray*}
c_{11} &=&2[\frac{k_{3\downarrow }-q_{1}q_{2}k_{1\uparrow }}{kq_{2}}\cos
(kd)-\frac{q_{1}\mu _{2}k_{1\uparrow }-\mu _{1}k_{3\downarrow }}{k^{2}q_{2}}%
\sin (kd)],\text{ } \\
c_{12} &=&2[(\frac{k_{3\downarrow }k_{1\uparrow }q_{1}+\mu _{1}\mu _{2}}{%
k^{2}q_{2}}-1)\sin (kd)+\frac{\mu _{1}q_{2}+\mu _{2}}{kq_{2}}\cos (kd)],
\end{eqnarray*}

\begin{eqnarray*}
c_{21} &=&-2[\frac{k_{3\downarrow }+q_{1}q_{2}k_{1\uparrow }}{kq_{2}}\cos
(kd)+\frac{\mu _{2}k_{1\uparrow }q_{1}+\mu _{1}k_{3\downarrow }}{k^{2}q_{2}}%
\sin (kd)], \\
\text{ }c_{22} &=&2[(\frac{k_{3\downarrow }k_{1\uparrow }q_{1}-\mu _{1}\mu
_{2}}{k^{2}q_{2}}+1)\sin (kd)-\frac{\mu _{1}q_{2}+\mu _{2}}{kq_{2}}\cos
(kd)],
\end{eqnarray*}

\begin{eqnarray*}
c_{31} &=&2[\frac{k_{3\downarrow }+q_{1}q_{2}k_{1\downarrow }}{kq_{2}}\cos
(kd)+\frac{\mu _{2}k_{1\downarrow }q_{1}+\mu _{1}k_{3\downarrow }}{k^{2}q_{2}%
}\sin (kd)],\text{ } \\
c_{32} &=&-2[(\frac{k_{3\downarrow }k_{1\downarrow }q_{1}-\mu _{1}\mu _{2}}{%
k^{2}q_{2}}+1)\sin (kd)-\frac{\mu _{1}q_{2}+\mu _{2}}{kq_{2}}\cos (kd)],
\end{eqnarray*}

\begin{eqnarray*}
c_{41} &=&2[\frac{k_{3\uparrow }-q_{1}q_{2}k_{1\uparrow }}{kq_{2}}\cos (kd)-%
\frac{q_{1}\mu _{2}k_{1\uparrow }-\mu _{1}k_{3\uparrow }}{k^{2}q_{2}}\sin
(kd)],\text{ } \\
c_{42} &=&2[(\frac{k_{3\uparrow }k_{1\uparrow }q_{1}+\mu _{1}\mu _{2}}{%
k^{2}q_{2}}-1)\sin (kd)+\frac{\mu _{1}q_{2}+\mu _{2}}{kq_{2}}\cos (kd)],
\end{eqnarray*}

\begin{eqnarray*}
c_{51} &=&-2[\frac{k_{3\uparrow }+q_{1}q_{2}k_{1\uparrow }}{kq_{2}}\cos (kd)+%
\frac{\mu _{2}k_{1\uparrow }q_{1}+\mu _{1}k_{3\uparrow }}{k^{2}q_{2}}\sin
(kd)], \\
\text{ }c_{52} &=&2[(\frac{k_{3\uparrow }k_{1\uparrow }q_{1}-\mu _{1}\mu _{2}%
}{k^{2}q_{2}}+1)\sin (kd)-\frac{\mu _{1}q_{2}+\mu _{2}}{kq_{2}}\cos (kd)],
\end{eqnarray*}

\begin{eqnarray*}
c_{61} &=&2[\frac{k_{3\uparrow }+q_{1}q_{2}k_{1\downarrow }}{kq_{2}}\cos
(kd)+\frac{\mu _{2}k_{1\downarrow }q_{1}+\mu _{1}k_{3\uparrow }}{k^{2}q_{2}}%
\sin (kd)],\text{ } \\
c_{62} &=&-2[(\frac{k_{3\uparrow }k_{1\downarrow }q_{1}-\mu _{1}\mu _{2}}{%
k^{2}q_{2}}+1)\sin (kd)-\frac{\mu _{1}q_{2}+\mu _{2}}{kq_{2}}\cos (kd)],
\end{eqnarray*}

\begin{eqnarray*}
a_{1} &=&\frac{q_{1}k_{1\uparrow }+i\mu _{1}}{k},\text{ }a_{2}=\frac{%
q_{1}k_{1\downarrow }+i\mu _{1}}{k}, \\
a_{3} &=&\frac{k_{3\uparrow }+i\mu _{2}}{kq_{2}},\text{ }a_{4}=\frac{%
k_{3\downarrow }+i\mu _{2}}{kq_{2}}.
\end{eqnarray*}

\bigskip {\bf FIGURE CAPTIONS}

Fig.1 The schematic layout of the F-N-F hybrid junction.

Fig.2 The relative orientation ($\theta $) dependence of the conductance for
different $\gamma $ [in atomic unit (au)], where $E_{F}=2.7eV$, $|{\bf h}%
_{L}|=|{\bf h}_{R}|=1.9eV$, $V_{1}=V_{2}=0.06$ $eV$, $\hbar =1,$ $%
m=m_{1}^{\ast }=m_{2}^{\ast }=m_{3}^{\ast }=1$ and $d=30$ \r{A}.

Fig.3 The $\gamma $-dependence of the conductance $G$ for different Fermi
energies, where $\theta =\pi /3$. The other parameters are chosen the same
as in Fig.2.

Fig.4 The $\gamma $-dependence of the $JMR$ for different Fermi energies,
where the parameters are chosen the same as in Fig.2.

Fig.5 $JMR$ as a function of $E_{F}$ for different $\gamma $'s. The
parameters are chosen the same as in Fig.2.

Fig.6 The thickness dependence of the $JMR$ for different $\gamma $'s. The
other parameters are chosen the same as in Fig.2.

Fig.7 (a) The $\gamma $-dependence of the conductance $G$ for different
effective masses; (b) The $E_{F}$-dependence of the conductance $G$ for
different effective masses, here $\gamma =0$. Both $\theta =\pi /3,$ $%
m_{2}^{\ast }=1$, $d=50$ \r{A}, and the other parameters are chosen the same
as in Fig. 2.

Fig.8 The $m_{3}^{\ast }$-dependence of the $JMR$ for different $m_{1}^{\ast
}$'s, here $m_{2}^{\ast }=1$ and $\gamma =0$. Inset: The contour plot for
the maximum $JMR$ in $m_{1}^{\ast }$-$m_{3}^{\ast }$ plane. The other
parameters are chosen the same as in Fig. 2.

Fig.9 The thickness dependence of the $JMR$ for different effective masses,
where $m_{2}^{\ast }=1$,  $\gamma =0$, and the other parameters are chosen
the same as in Fig.2.

Fig.10 The $\gamma $-dependence of the $JMR$ for different $m_{3}^{\ast }$%
's, where $m_{2}^{\ast }=1$, and the other parameters are chosen the same as
in Fig.2. (a) $m_{1}^{\ast }=0.05$; (b) $m_{1}^{\ast }=0.5$; (c) $%
m_{1}^{\ast }=5$.

\end{document}